\documentclass[10pt,letterpaper]{article}

\usepackage{changepage}

\usepackage[utf8]{inputenc}

\usepackage{textcomp,marvosym}

\usepackage{fixltx2e}

\usepackage{amsmath,amssymb}

\usepackage{cite}

\usepackage{nameref,hyperref}

\usepackage[right]{lineno}

\usepackage{microtype}
\DisableLigatures[f]{encoding = *, family = * }

\usepackage{rotating}

\usepackage[aboveskip=1pt,labelfont=bf,labelsep=period,justification=raggedright,singlelinecheck=off]{caption}

\newcommand{\beq}[1]{\vspace{-0.02in}\begin{equation}#1\end{equation}\vspace{-0.02in}}
\newcommand{\besp}[1]{\begin{split}#1\end{split}}

\date{}

\usepackage{lastpage,fancyhdr,graphicx}
\usepackage{epstopdf}
\begin{document}
\vspace*{0.35in}

\begin{flushleft}
{\Large
\textbf\newline{Inferring Social Status and Rich Club Effects in Enterprise Communication Networks}
}
\newline
\\
Yuxiao Dong\textsuperscript{1},
Jie Tang\textsuperscript{2},
Nitesh V. Chawla\textsuperscript{1},
Tiancheng Lou\textsuperscript{3},
Yang Yang\textsuperscript{1},
Bai Wang\textsuperscript{4}
\\
\bigskip
{\bf 1} Interdisciplinary Center for Network Science and Applications, Department of Computer Science and Engineering, University of Notre Dame, Notre Dame, IN, United States of America
\\
{\bf 2} Department of Computer Science and Technology, Tsinghua University, Beijing, P. R. China
\\
{\bf 3} Google Inc, Mountain View, CA, United States of America
\\
{\bf 4} Department of Computer Science and Technology, Beijing University of Posts and Telecommunications, Beijing, P. R. China
\\
\bigskip

%
%





* nchawla@nd.edu

\end{flushleft}
\section*{Abstract}
Social status, defined as the relative rank or position that an individual holds in a social hierarchy, is known to be among the most important motivating forces in social behaviors.  In this paper, we consider the notion of status from the perspective of a position or title held by a person in an enterprise. We study the intersection of social status and social networks in an enterprise. We study whether enterprise communication logs can help reveal how social interactions and individual status manifest themselves in social networks.
To that end, we use two enterprise datasets with three communication channels --- voice call, short message, and email --- to demonstrate the social-behavioral differences among individuals with different status. We have several interesting findings and based on these findings we also develop a model to predict social status.
On the individual level, high-status individuals are more likely to be spanned as structural holes by linking to people in parts of the enterprise networks that are otherwise not well connected to one another.
On the community level, the principle of homophily, social balance and clique theory generally indicate a ``rich club'' maintained by high-status individuals, in the sense that this community is much more connected, balanced and dense.
Our model can predict social status of individuals with 93\% accuracy.


\section*{Introduction}
Social status refers to the relative rank (or position) that an individual holds in a society~\cite{Rege:08,Kleinberg:10BOOK}.
For instance, an indicator of status in an enterprise setting could be the position that an individual holds in the company; and an indicator of status in a social media (such as Twitter, Facebook, etc.) could be the number of `followers' or `friends' one has.
Apart from economic incentive, social status is known to be among the important motivating forces in social behaviors~\cite{SocalClass:92}, and achieving a higher status in a social network is often a motivating force in influencing an individual's behavior~\cite{Status64:Weber,Rodkin:2008}.
An individual with higher status is likely to have more influence or control on the message, mission, or vision within a social system. 

However, the relationship between social network structure and social status has not been widely recognized or studied, especially from the perspective of communication channels, such as voice call, short message, and email. 
Researchers have studied the interplay of social status and language signals~\cite{Danescu:WWW2012,Gilbert:CSCW12}, cognitive politics~\cite{status:14PLoS}, management science~\cite{Tichy:1979}, race perception~\cite{status:11PLoS} and collectivist cultures~\cite{Cecilia:CSCW14}. 
Csermely et al.~\cite{CorePeriphery:Uzzi13} find that the core and periphery network structure plays an important role in the development of a large variety of complex organisms and organizations.
Tang et al.~\cite{Tang:12WSDM,Dong:ICDM12,Lou:TKDD13} illustrate how the opinion leaders and ordinary users correlate to social tie formation and type in social behaviors.
Leskovec et al.~\cite{Leskovec:10CHI,Leskovec:10WWW} develop a theory of status in online social networks, which provides a different organizing principle for directed networks of signed links. In addition, researchers have inferred demographic information such as age or gender from social networks~\cite{Dong:KDD14,Bara:Sex12,sr13:women,PNAS13:motif}.
Essentially, these studies derive an individual's status from computing node centralities or other node attributes such as demographics, rather than the social status of the nodes. 
In this paper, we consider social status from the perspective of position or title in an organization and its inter-relationship with social networks defined by communication channels. 

The dynamics of social status can actively influence the strategies to make friends and maintain connections, which could re-structure the social circles of individuals. The social relationships of individuals can also rearrange the promotion or demotion of their social status and further reform the social hierarchy (or circles) in the organization. 
Furthermore, social status might also impact how information diffuses or cascades in a network via different communication channels. 
We discover that there are, indeed, different communications and social networking underpinnings as a consequence of individuals with different status in a social network. This can be especially compelling in a corporate organization that by design has individuals with different positions (or status) --- managers and their direct reports (or subordinates).

To that end, we collected three communication networks from two different enterprises.
Two mobile communication networks, i.e., voice call (CALL) and short message (SMS) networks, are extracted from two-month communication logs of an Asian telecommunication company with 50 managers and 182 subordinates. 
And an email communication (EMAIL) network is from Enron Inc. with 155 managers and 22,322 subordinates spanning over one year~\cite{enron}.  This allows us to consider nuances of not only the two different enterprises but also three different communication channels. 
Specifically, we posit and evaluate whether communication patterns in an organization vary depending on the social status of an individual, and inversely whether we can effectively build a model to infer social status using the communication patterns. 
We indicate the relative ranks of users in the company as their social status, i.e., managers (M) as high status and subordinates (S) as low status (Please see \textit{Materials and Methods} for additional details about the data).  

We consider the following questions in this paper: 
1) What are the fundamental clues or patterns that may subtly reveal individuals' status in social networks?
2) How does the status of individuals influence the formation of network structure?
3) How accurately can we infer individuals' status from social network structure? 

We observed that across the different communication channels (CALL/SMS/EMAIL), there exist consistent patterns arising from social status and its impact on network structure. We also discovered that various social theories and characteristics can be indicative of social status. The key findings, validated at a 95\% statistical significance, include: 
\begin{itemize}
\item High-status individuals are more likely to be spanned as ``structural holes'' in networks than their  subordinates, which indicates that she or he is linked to individuals in different parts of the network that are otherwise not well connected to one another~\cite{Burt:95}. Thus the ``managers'' act as the bridge across groups. 
\item At the neighborhood level, the number of common connections maintained by high-status individuals is over three times higher than the number of common connections maintained by the subordinates.
\item The social community among high-status users is much more balanced and denser than subordinates, which further unveils the ``rich club''~\cite{Vespignani:NaturePhys2006} effects of  high-status employees in enterprise networks. 
\end{itemize}

Inspired by the observations around the social structure and characteristics, and their potential to infer social status in a network, we also developed a probabilistic graphical model to predict social status using the aforementioned characteristics as features.  The proposed model, referred to as Factor Graph Model (FGM), associates a latent variable to each user in the communication network to represent her or his social status. We demonstrate that the presented model can accurately infer as many as 93\% of social users' status by leveraging the correlations between network structure and social status.

\section*{Results}
\subsection*{Communication Behaviors}

We first examine the communication patterns of the different staff (managers and subordinates) in the enterprise. Note, we are only focusing on the intra-company communication behaviors. 

\begin{figure*}[!ht]
\begin{center}
\includegraphics[width=6in]{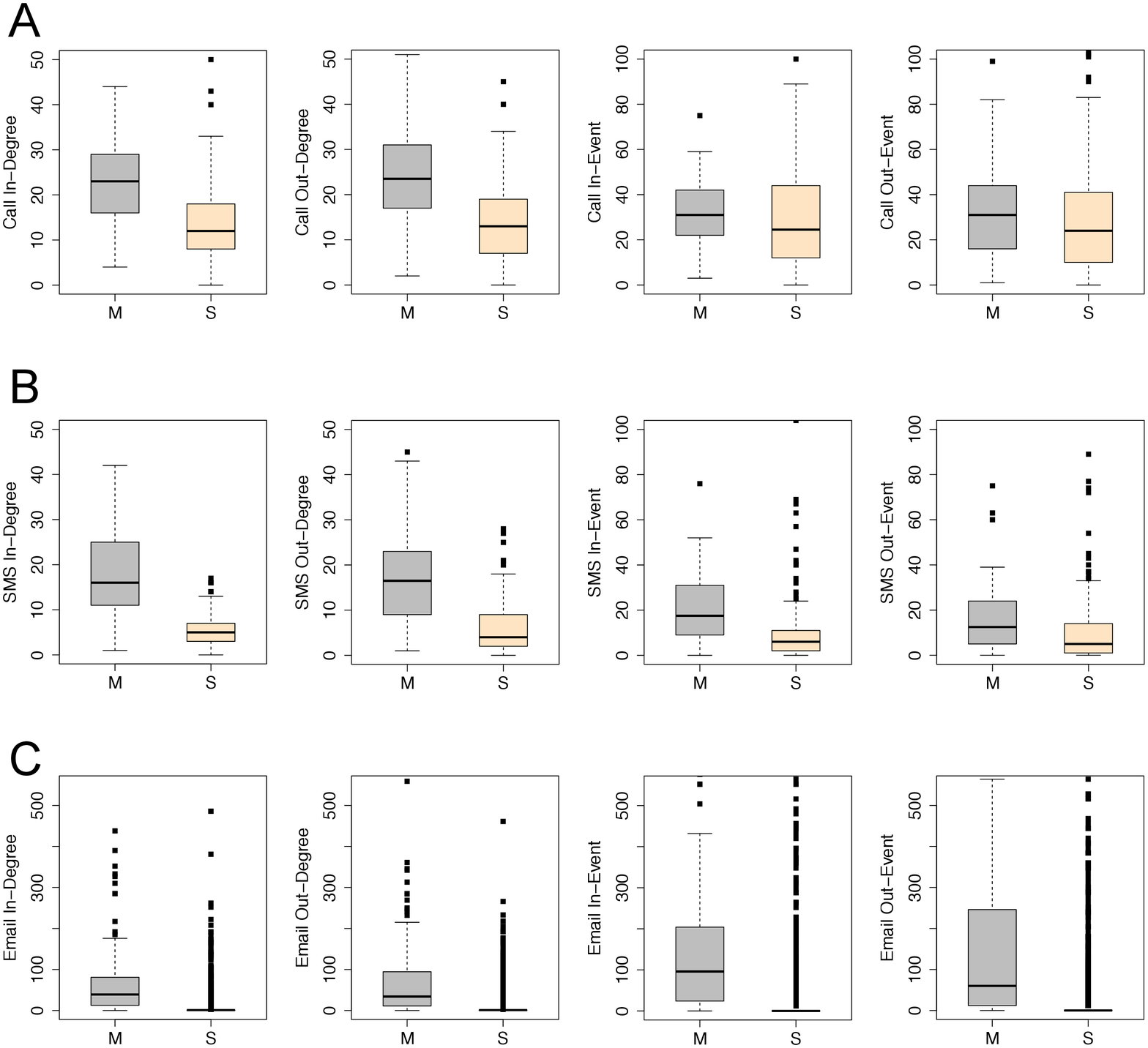}
\end{center}
\caption{\label{fig:comm-charac}
{\bf Communication Attributes vs. Social Status.} (A). CALL attributes; (B). SMS attributes; (C). EMAIL attributes. Each box plot represents the distribution of the number of attributes per time unit (each month in A and B, and each year in C), including in-degree, out-degree, in-event and out-event.
}
\end{figure*}

Fig. \ref{fig:comm-charac} shows the differences of four communication characteristics, including in-degree, out-degree, in-event, and out-event, between managers and subordinates from the three different channels --- CALL, SMS, and EMAIL --- across the two different companies. For each channel, one's in- or out- degree is defined by the number of contacts who make or receive the communications, and the number of events is defined by the count of communications. 
We find that managers use mobile phones or emails more frequently than their subordinates. For example, on average, each manager makes about 60 calls (out-event) to 25 receivers (out-degree) in two months, while each subordinate only makes 40 calls to about 10 people.
We can also see that both managers' number of calls and number of receivers are around three to four times that of a subordinate. We also compare the difference among various communication channels.
Clearly, there is a larger gap between manager and subordinate in SMS than CALL behaviors.
As for EMAIL, the communication differences on frequency between managers and subordinates become more substantial than the mobile channels.
Thus, the characteristics of interaction across different communication channels have the potential to reveal the social status of individuals in networked communications. 
We also studied duration of phone calls between different staff but found no significant difference between managers and subordinates in the actual duration of CALLs.

\subsection*{Social Status vs. Social Theories}

Social status and the resulting patterns of communication characteristics, across the three different networks, give us an opportunity to study the interplay between individuals' status and their network influence via several well-known social theories, including structural hole, social balance, homophily and social clique.
The observations from CALL, SMS and EMAIL channels unveil strong ``rich club'' phenomenon~\cite{IEEE2004:RichClub,Julian:APL2007} in enterprise networks.
First, managers are much more likely to be spanned as ``structural holes'' in networks than subordinates.
Second, they also maintain 3-8 times more common neighbors compared to subordinates.
Third, the managers form more balanced triads than those by the subordinates. 
Finally, the social community among managers is much closer and denser than subordinates. 
We use a null model on the randomized version of the empirical data and report the statistical significance of the results on structural hole and social balance. 
The details of the null model are introduced in \textit{Materials and Methods}. \\

\noindent\textbf{Structural Hole.} The principle, that individuals can benefit from serving as intermediaries between others who are not directly connected, forms the underpinning for the theory of structural holes \cite{Burt:95,Burt04:hole,Jon:2008EC,Tang:WWW13}.
For example, a manager in a department with a diverse range of connections can be considered as a structural hole, with a number of ties to staff in other different departments.
The fundamental question is: do high-status individuals span as structural holes in networks?
Here we consider the HIS algorithm \cite{Tang:WWW13} to estimate the likelihood of each node in the network to span as structural hole, and then categorize them into two groups based on the likelihood.
We select the top 21\% (the percentage of managers in mobile networks) of individuals in CALL and SMS networks and the top 0.67\% individuals (the percentage of managers in EMAIL network) in EMAIL Network as structural holes based on their HIS scores and the rest as ordinary individuals.
Our conjecture is that if the status does not correlate with structural holes, the probability that managers are structural holes should be the same as the ratio of managers (21\% in mobile and 0.67\% in EMAIL). 
However, our analysis in Table \ref{tb:sh_pvalue} clearly shows that managers are more likely (70\% in CALL, 55\% in SMS, and 43\% in EMAIL) to be spanned as structural holes across the three networks.
In other words, the structural holes extracted from enterprise communication network structure reveal the social status of staff in their company.
This can be explained by the fact that managers usually need to operate the responsibility of correspondents and organizers within the company, especially for the experience for connecting different departments or groups to cooperate. \\

\begin{table}[!ht]
\centering
\caption{{\bf Structural Holes vs. Social Status.}
The probability that staffs in companies are spanned as structural holes (SH) extracted from communication network structures. M: Managers; S: Subordinates. ($* p<0.05; ** p<0.01; *** p<0.001; **** p<0.0001$)
}
\begin{tabular}{|l|l|l|l|}
\hline
                 & CALL         &  SMS        & EMAIL  \\ \hline
M as SH          & 0.700 ****    &  0.550 ****  & 0.430 ****    \\ \hline
M as SH (Random) & 0.207         & 0.207        & 0.007  \\ \hline
S as SH          & 0.300 ****    &  0.450 ****  & 0.570 ****   \\ \hline
S as SH (Random) & 0.793         & 0.793        & 0.993    \\ \hline
\end{tabular}
\label{tb:sh_pvalue}
\end{table}

\begin{figure}[!ht]
\begin{center}
\includegraphics[width=3.27in]{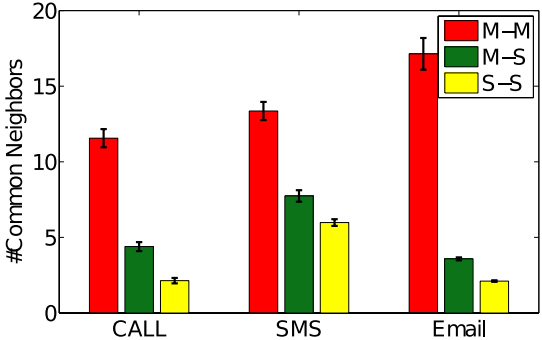}
\end{center}
\caption{
{\bf Homophily vs. Social Status.} The number of common friends of each pair of employees are considered as the measurement of link homophily in social network, thus generating three different types of ties: M-M, M-S, S-S. Error bars show the confidence interval with significance level $\alpha=0.05$.
}
\label{fig:homo}
\end{figure}

\noindent\textbf{Link Homophily.} Homophily is the tendency of individuals to associate and bond with similar others \cite{Lazarsfeld:1954homophily,McPherson:01}.
The presence of homophily has been widely discovered in some form or another, including age, gender, class, and organizational role.
Lazarsfeld and Merton \cite{Lazarsfeld:1954homophily} argued that individuals with similar social status are more likely to associate with each other, which is called status homophily. 
Particularly, following the theory of homophily~\cite{McPherson:01}, we consider the neighbors of one individual as her or his attributes and examine the correlation of the neighbors of different individuals. Then, the concept of link homophily~\cite{Lou:TKDD13} tests whether two individuals who share more common neighbors will have a tendency to have similar social status in the company. 
The average number of common neighbors by two managers ranges from 12 to 17 across the three networks from Fig. \ref{fig:homo}.
Surprisingly, the average number of common neighbors of pairs of subordinates only reaches around two in CALL or EMAIL networks and six in SMS network.
As homophily phenomena gets more reflected among managers, we can contend that two individuals are much more likely to be two managers in the company if they share more common neighbors.
Managers' ability of creating and maintaining social connections in enterprise networks is more prominent than subordinates'.
This could have the potential to further promote their status in companies, which further highlights the rich club effect. \\

\begin{figure}[!ht]
\begin{center}
\includegraphics[width=3.27in]{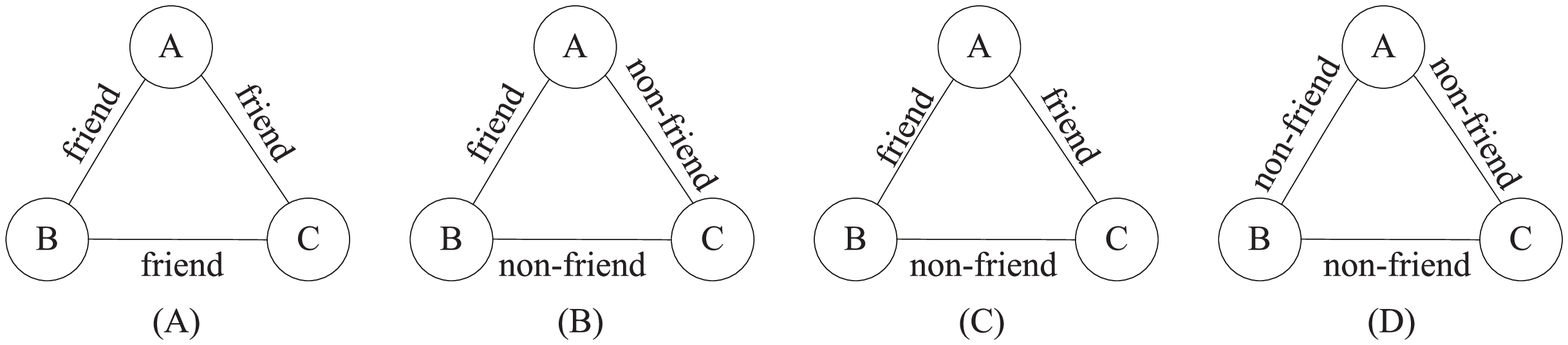}
\end{center}
\caption{\label{fig:sb_wsdm}
{\bf Illustration of structural balance theory.} Triads (A) and (B) are balanced, while (C) and (D) are not balanced.}
\end{figure}

\noindent\textbf{Social Balance.} Triad is one of the simplest grouping of individuals that can be studied and is mostly investigated by microsociology \cite{Kleinberg:10BOOK}.
Given a triad $(A, B, C)$, if $A$ and $B$ are friends and if $A$ and $C$ are friends as well, then social balance theory~\cite{Balance:Heider1958} suggests that $B$ and $C$ are also likely to be friends, which results in a balanced triad. Similarly it is also a balanced triad where $A$ and $B$ are friends, while  $B$ and $C$, and $A$ and $C$ are non-friends. The other structured triads are called unbalanced triads. 
For each node, the balance ratio is defined as the ratio of the number of balanced triads to the number of unbalanced triads. 
An illustrative example is shown in Fig. \ref{fig:sb_wsdm}.
According to the social balance theory, a stable social network tends to be a balanced structure by containing densely connected triads~\cite{Balance:PLOS2013}.
Here, we aim to test whether the enterprise communication networks also have balanced structures with respect to social status.
Given one individual and her/his ego network, we calculate three kinds of social balance ratios, i.e., the balance ratio among her/his manager-friends M-$sb$, the balance ratio among her/his subordinate-friends S-$sb$, and the overall balance ratio among all her/his friends $sb$.
We find that the managers' overall balance ratios are larger than the subordinates' across all the three channels in Table~\ref{tb:balance_pvalue}.
Moreover, the managers are more likely to form balanced structure among their manager-friends, and the subordinates with subordinates.
In other words, the individuals in organizations have the tendency to create or maintain balanced relationships with people of the same status; this  phenomenon coincides with the link homophily observed above.
We conjecture that the relatively high status empowers the managers to connect with more people and maintain the relationships within the enterprise, enhancing the chance to promote their status. \\

\begin{table}[!ht]
\centering
\caption{{\bf Social Balance vs. Social Status.}
The social balance ratio of staffs with different status in companies in three communication networks. M: Managers; S: Subordinates. M-$sb$: social balance ratio of manager-friends; S-$sb$: social balance ratio of subordinate-friends; $sb$: social balance ratio of all friends.
($* p<0.05; ** p<0.01; *** p<0.001; **** p<0.0001$)
}
\begin{tabular}{|l|l|l|l|l|l|l|}
\hline
                    & M (CALL)  & S (CALL)  & M (SMS)   & S (SMS)   & M (EMAIL) & S (EMAIL) \\ \hline
M-$sb$              & 0.569**** & 0.348**** & 0.546**** & 0.468****  & 0.455****& 0.047****     \\ \hline
S-$sb$              & 0.174***  & 0.254**   & 0.289     & 0.299   & 0.066**     & 0.082****   \\ \hline
$sb$                & 0.340     & 0.312     & 0.325     & 0.311   & 0.165     & 0.124     \\ \hline

\end{tabular}
\label{tb:balance_pvalue}
\end{table}

\begin{figure}[!ht]
\begin{center}
\includegraphics[width=3.27in]{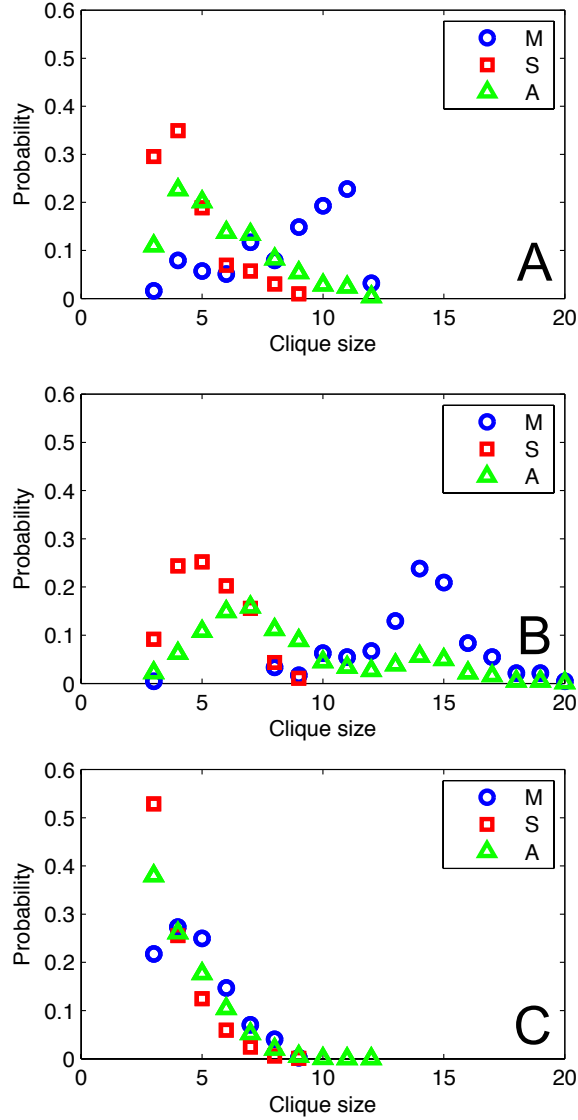}
\end{center}
\caption{\label{fig:clique}
{\bf Social Clique vs. Social Status.} Distribution of social clique in enterprise communication networks. M: Managers; S: Subordinates; A: All employees.
}
\end{figure}

\noindent\textbf{Social Clique.} Clique is a concept in both social sciences and graph theory. In social sciences, clique is used to describe a group of persons who interact with each other more regularly and intensely than others in the same setting~\cite{Clique:2008BOOK}. 
In graph theory, a clique is defined as a subset of nodes such that for any two nodes, there exists an edge connecting them~\cite{Clique:1973}.
Interacting with people in one clique has the indication of close and strong relationships with each other. 
Here we aim to examine how managers and subordinates form cliques and to which extent they are connected.
We build two sub-networks that only contain mangers or subordinates respectively for each type of a network derived from each of the communication channels.
Fig. \ref{fig:clique} shows the distributions of clique size, conditioned on the status of individuals (employees in the enterprise). 
For reference, we also plot the overall clique distribution in each full network.
It is obvious that the distributions of managers and subordinates are quite different. The maximal cliques for 50 managers are 12, 20 in CALL and SMS networks, respectively. It is interesting that the clique sizes vary across the two different communication channels, albeit in the same company. The clique size for the 155 managers in EMAIL is 9. In comparison, the maximal clique sizes for 182 sub-ordinates in the CALL and SMS networks are 9 and 10, respectively; and the clique size for the 22,232 subordinates in the EMAIL network is 9. 
We also find that the most frequent cliques in subordinates' sub-networks are 4/5/3-clique in CALL/SMS/EMAIL, respectively, which are much smaller than the 11/13/4-clique in the CALL/SMS/EMAIL for managers' sub-networks. 

In addition, we note the difference in the network properties between the two sub-networks in Tables \ref{tb:table_subnetTopo}, even though they belong to the same enterprise mobile networks. 
Specifically, the clustering coefficient for sub-network (M) is two times higher than sub-network (S), which coincides with our observations above that the managers share more dense connections than the sub-ordinates in an organization. 
The correlations between social status and several social theories provide the evidence of ``rich club'' maintained by high-status individuals.

\begin{table}[!ht]
\centering
\caption{{\bf Topology characteristics of sub-networks.}
M: Managers; S: Subordinates.
}
\begin{tabular}{|l|r|r|r|r|r|r|}
\hline
attributes                              & M(CALL) & S(CALL)  & M(SMS) & S(SMS)  & M(EMAIL) & S(EMAIL) \\ \hline
{\#nodes}                                & 50 & 182           & 50 & 182         & 155 & 22322  \\ \hline
{\#edges}                                 & 511 & 1858       & 693 & 1627        & 964 & 29912  \\ \hline
{clustering coefficient}              & 0.63 & 0.31       & 0.78 & 0.45      & 0.46 & 0.10  \\ \hline
{associative coefficient}              & -0.04 & 0.02      & -0.29 & -0.12   & 0.03 & -0.14    \\ \hline
\end{tabular}
\label{tb:table_subnetTopo}
\end{table}

\subsection*{Social Status Prediction}

We now consider the core problem: can we leverage the observations from communication behaviors to infer an individual's social status? 
We use several classical data mining models to infer social status, including Naive Bayes (NB), Bayes Network (BNET), Logistic Regression Classification (LRC) and Conditional Random Fields (CRF) \cite{Lafferty:01}.
We also propose a Factor Graph Model (FGM) to leverage the social theories to help status prediction.
We use Weka~\cite{Weka:KDDEXP09} for NB, BNET and LRC methods. NB, BNET and LRC use communication attributes to train classification models and apply them to predict individual status. For CRF and FGM models, both communication and social features are used to infer the labels of individuals.
We quantitatively evaluate the performance of inferring individual status in terms of weighted Precision, Recall, F1-Measure and Accuracy. 

Table \ref{tb:socialstatus} shows the results of four methods for inferring individual status in these communication networks. Clearly, our model FGM yields better performance than other alternative methods.
In CALL network, we can find that FGM achieves about 85\% F1-Measure scores and Accuracy. In the SMS network, both F1-Measure and Accuracy of FGM reach 92\%.
However, the text-messaging network seems to reveal more status differentials than the calling network.
By combining the CALL and SMS networks together, the predictive performance can be improved by 2\%-3\% compared with the results from the SMS network.
The prediction performance from EMAIL channel outperforms the other two mobile channels by 1-8\% in terms of Accuracy.
In summary, the social status of 85\% -- 93\% of individuals can be inferred from their communication interactions among their colleagues.
This prediction results further confirm our observations on communication behaviors and social theories are general across different companies, even with different communication channels (CALL vs. SMS vs. EMAIL).

\begin{table}[!ht]
\caption{{\bf Social status prediction in enterprise communication networks.}
NB: Naive Bayes, BNET: Bayes Network, LRC: Logistic Regression Classification, CRF: Conditional Random Fields, FGM: Factor Graph Model.
}
\label{tb:socialstatus}
\centering
\renewcommand\arraystretch{1.1}
\begin{tabular}{|c|c|c|c|c|c|}
\hline
Status & Method & Precision & Recall & F1 & Accuracy\\
\hline

CALL     & NB            & 0.7334 & 0.7625 & 0.7416 & 0.7625 \\ \hline
CALL     & BNET        & 0.7409 & 0.6934 & 0.7110 & 0.6934\\ \hline
CALL     & LRC         & 0.7065 & 0.6795 & 0.6904 & 0.6795\\ \hline
CALL     & CRF         & 0.8078 & 0.8095 & 0.8086 & 0.8095\\ \hline
CALL     & FGM & {0.8514} & {0.8508} & {0.8511} & {0.8508}  \\
\hline

SMS      & NB             & 0.8693 & 0.8734 & 0.8648 & 0.8734 \\ \hline
SMS      & BNET        & 0.8497 & 0.8512 & 0.8483 & 0.8512\\ \hline
SMS      & LRC         & 0.8129 & 0.7850 & 0.7935 & 0.7850\\ \hline
SMS      & CRF         & 0.8720 & 0.8761 & 0.8740 & 0.8760\\ \hline
SMS      & FGM & {0.9321} & {0.9276} & {0.9298} & {0.9276} \\
\hline

EMAIL    & NB          & 0.8847 & 0.8993 & 0.8847 & 0.8598 \\ \hline
EMAIL    & BNET        & 0.8936 & 0.9054 & 0.8164 & 0.8755\\ \hline
EMAIL    & LRC         & 0.8761 & 0.8772 & 0.7653 & 0.8483\\ \hline
EMAIL    & CRF         & 0.9033 & 0.8902 & 0.8967 & 0.8902\\ \hline
EMAIL    & FGM & {0.9319} & {0.9383} & {0.9373} & {0.9383}  \\

\hline

\end{tabular}
\end{table}

\section*{Discussion}
Interactions within a corporate enterprise are representative of the artifacts of the status of individuals in the enterprise.  To that end, we analyzed the communication interactions (call, message, email) of managers and subordinates in a corporation by network properties.
We find that the managers (or the high-status individuals) in a corporate hierarchy congregate as a ``rich club'', maintaining denser and closer connections than the subordinates (or the low-status individuals) in the same enterprise. 
This phenomenon stands out with different social characteristics across the networks derived from different communication channels. This result also becomes evident from the larger social cliques of the managers. 

The structure of communication networks collected from different channels also suggests that the managers (high-status) are more likely to be spanned as structural holes and maintain more balanced social circles than the subordinates (low-status).
The number of common neighbors also indicates the stronger homophily between the high-status individuals. 
The social circles maintained by high-status individuals are much larger and closer than subordinates, simultaneously, the social capital, namely the collective benefits from social circles, that managers own are much greater than subordinates based on their social circles. 

Finally, we study to what extent the social status of individuals can be inferred from their communication network structure. That is, can the observed communication and social characteristics in networks be used to develop a model for inferring the actual social status? This is an important experiment for lending an insight into predicting the status of an individual when the only observed information is the social network and the patterns of communication behavior. 
We propose a factor-graph based model, and demonstrate that our model is able to achieve about 85\% of predictive accuracy using CALL network, about 92\% of predictive accuracy using SMS, and over 93\% of predictive accuracy using EMAIL. 
The performance trends clearly show that we are able to capture essential properties of social theories, which are general across different communication channels in different companies.

Social status characterizes the strategies that people organize their social connections. 
It offers a great potential to understand the underlying principles that drive human social activities and behaviors. 
Awareness of individual status can provide new perspectives in network science problems, such as link prediction, influence propagation, and community detection, which were considered in black-and-white network structure before.

\section*{Materials and Methods}

\subsection*{Communication Network Data}

The mobile dataset used in this paper is extracted from a large collection of call and text-message records, which span over two months.
We construct mobile communication networks for all employees in a telecommunication company with 15 departments, where there are 232 staff, which include one CEO, four Vice Presidents, 45 department managers (each department has three managers) and 182 subordinates.
However, the dataset provided for the study contains only two levels of status (manager or subordinate) for each individual, resulting into two groups of managers (high-status) and subordinates (low-status). We construct two sub-networks for the mobile enterprise --- one using the voice calls (CALL) and the other using the text messaging service (SMS). 

The EMAIL network is extracted from the Email communication logs of Enron Inc. \cite{enron,EnronOnline}.
It consists of 164,080 emails between 22,477 Enron employees, including 155 managers and 22,322 subordinates.
Table \ref{tb:table_netTopo} lists statistics of the three networks. $cc$ is the average clustering coefficient, $ac$ is the associative coefficient and $cn$ denotes the number of components in the network.

\begin{table}[!ht]
\centering
\caption{{\bf Network topology characteristics.}
M: Managers; S: Subordinate.
}
\begin{tabular}{|l|r|r|r|r|}
\hline
attributes                  & CALL    & SMS          &  EMAIL \\ \hline
{\#nodes}              & 232               & 232     &  22477  \\ \hline
{\#edges}              & 3340              & 3406    &  44728 \\ \hline
{clustering coefficient}          & 0.3326  & 0.4761 &  0.1241\\ \hline
{associative coefficient}      & 0.1195  & -0.0894   & -0.2153 \\ \hline
\end{tabular}
\label{tb:table_netTopo}
\end{table}

\subsection*{Null Model for Different Social Status of Staff}
We use a null model~\cite{PNAS13:motif} to validate the statistical significance of our social observations.
A straightforward way to measure this is to compare the real values to the null model where the status of people is randomly assigned.
We compare the real data to 10,000 randomized cases where managers and subordinates are randomly shuffled.
First, we simulate the random process of allocating status to individuals with the same ratio as in the real data (50 managers and 182 subordinates of mobile networks, and 155 managers and 22,322 subordinates of EMAIL network) 10,000 times for the underlying network structure.
The difference on social observation between empirical data $x$ and the null model $\tilde{x}$ can provide the interpretation for the deviation. 
The $z$ score can examine whether the null model is true, i.e., there are no distinctions between individual status given the underlying communication network structure.
$$z(x) = \frac{x - \mu(\tilde{x})}{\sigma(\tilde{x})}$$  
\noindent where $\mu(\tilde{x})$ and $\sigma(\tilde{x})$ are the mean and standard deviation of the observations on the null model. 
The null hypothesis is rejected at 2 sigmas (corresponding to the  $p$-value $<$ 0.01).

\subsection*{Factor Graph Model for Status Prediction}

From the machine learning perspective, if we consider each individual as an instance in a learning model, we will speak of each individual as a relatively ranking position (such as manager and subordinate).
If we assume that all data instances (individual-based instances) are independent, then we can leverage standard machine learning algorithm to learn a classifier. 
However, the instances are not necessarily independent. To that end, we also leverage a  probabilistic graphical model~\cite{Graphical:FTML08} to build our models. We consider factor graph~\cite{Kschischang:01} that is able to model the correlations among variables. We have also used The Factor Graph Model (FGM) in our previous works~\cite{Lou:TKDD13,Dong:KDD14}.

Let $G = (V, E)$ denote the communication network, where $V$ is a set of individuals and $E\subseteq V \times V$ is a set of edges. Each edge $e_{uv}$ is created if individual $u\in V$ and individual $v\in V$ have communication logs between each other.
In the FGM model, the communication network $G$ is directly transformed as a factor graph with each node as an individual and each edge as communication relationship between two individuals. For each individual node $v\in V$, a hidden variable $y_i$ is introduced to represent the relative rank (social status) of the corresponding individual.
For example, in our mobile network, we use each individual's position in a mobile company as the status, thus we can define two ranks for $y$ to respectively represent manager and employee.
Given some labeled training data $(G, \textbf{X}, Y)$, where $\textbf{X}$ is the individual attribute matrix, the objective function can be defined as a log-likelihood function
$$\mathcal{O}(\theta)=\log P(Y|G, \textbf{X}, \theta)$$

\noindent where $\theta$ are parameters to learn from the training data.
If we consider $P(.)$ as an exponential distribution over various available features in the social network, we can formally define the log-likelihood objective function as:

\beq{\label{eq:objN}
\besp{
\mathcal{O}(\theta) = \sum_{v}^V \sum_k^K \theta_k f_k(x_v, y_v) +  \sum_{c}^C \theta_c f_c(y_u, y_v, y_w) -\log Z
}
}

\noindent where $f_k(x_v, y_v)$ is the $k$-th feature defined over node $v$, $K$ is the number of features for each node, $f_c(y_u, y_v, y_w)$ is the $c$-th correlation feature defined over each triangle $c = \{u, v, w| e_{uv}, e_{uw}, e_{vw} \in E\}$, $C$ is the set of all closed triangles in the graph $G$, and $Z$ is a normalization factor. 

For model learning, the task is to find a parameter configuration $\{\theta\}$ to maximize the log-likelihood objective function
Eqs. \ref{eq:objN}, i.e.,

\beq{
    \theta^* = \arg \max \mathcal{O} (\theta)
}

\noindent In this work, we use a gradient descent method (or a Newton-Raphson method) to optimize the convex objective function~\cite{Graphical:FTML08,Lafferty:01}. Specifically, we first write the gradient of each $\theta_k$ with regard to the objective function:

\beq{ \label{eq:gradient}
    \frac{\partial \mathcal{O}(\theta)}{\partial \theta_k} = \mathbb{E}[f_k(x_v, y_v)] - \mathbb{E}_{P_{\theta_k}(y_{v}|x_{v}, G)}[f_k(x_v, y_v)]
}

\noindent where $\mathbb{E}[f_k(x_{v}, y_{v})]$ is the expectation of feature function $f_k(x_{v}, y_{v})$ given the data distribution and $\mathbb{E}_{P_{\theta_k}(y_{v}|x_{v}, G)}[f_k(x_{v}, y_{v})]$ is the expectation of feature function $f_k(x_{v}, y_{v})$ under the distribution $P_{\theta_k}(y_{v}|x_{v}, G)$ given by the estimated model.

The graphical structure in the above model can be arbitrary and may contain circles, which makes it intractable to directly calculate the marginal distribution $P_{\theta_k}(y_{v}|x_{v}, G)$. 
To solve this challenge, we use Loopy Belief Propagation, due to its ease of implementation and effectiveness, to approximate the marginal distribution $P_{\theta_k}(y_{v}|x_{v}, G)$. Then we are able to obtain the gradient by summing up all the factor graph nodes. Finally, we update each parameter with a learning rate $\eta$ with the gradient. Related algorithms can be found in \cite{Lou:TKDD13,Status:KDD13}.

With the estimated parameter $\theta$, we can now assign the value of unknown labels $Y$ by looking for a label configuration that will maximize the objective function, i.e.

\beq{\label{eq:Ymax}
    Y^* = \arg \max \ \mathcal{O}(Y|G, \textbf{X}, \theta)
}

\noindent Obtaining exact solution is again intractable. The LBP is utilized to calculate the marginal probability for each node in the factor graph. Finally, labels that produce the maximal probability will be assigned to each factor graph node.


\section*{Acknowledgments}
YD, NVC and YY are supported by the Army Research Laboratory and was accomplished under Cooperative Agreement Number W911NF-09-2-0053, the United States Air Force Office of Scientific Research and the Defense Advanced Research Projects Agency grant \#FA9550-12-1-0405; JT is supported by the National High-tech R \& D Program (No. 2014AA015103), National Basic Research Program of China (No. 2014CB340500, No. 2012CB316006), Natural Science Foundation of China (No. 61222212), a research fund supported by Huawei Inc. and Beijing key lab of networked multimedia; BW is supported by National Basic Research Program of China (2013CB329603).

%
%
%



\small
\bibliographystyle{abbrv}
\bibliography{references-full}

\end{document}